\documentclass[showpacs,prd,notitlepage,nofootinbib]{revtex4-1}
\usepackage[utf8]{inputenc}
 \usepackage{color,graphicx,epsfig}
 \usepackage{amsmath,amssymb,dsfont,epsfig,graphicx,xcolor,fontenc}
 \usepackage{ifpdf}
 \usepackage{amsmath}
 \usepackage{bm}
 \usepackage[english]{babel}
 \usepackage{graphicx}%
 \usepackage{amsfonts}%
 \usepackage{amssymb}
 \usepackage{braket}
 \usepackage{hyperref}
 \usepackage{epstopdf}
\newcommand{\change}[1]{#1}
\def\Argentina{International Center for Advanced Studies (ICAS) and CONICET, UNSAM,
Campus Miguelete, 25 de Mayo y Francia, CP1650, San Martin, Buenos Aires, Argentina}

\begin{document}

\title{Bayesian inference to study a signal with two or more decaying particles\\ in a non-resonant background}

\author{Ezequiel Alvarez}
\email[Electronic address:]{sequi@unsam.edu.ar}
\affiliation{\Argentina}

\begin{abstract}
We study the application of a Bayesian method to extract relevant information from data for the case of a signal consisting of two or more decaying particles and its background.  The method takes advantage of the dependence that exists in the distributions of the decaying products at the event-by-event level and processes the information for the whole sample to infer the mixture fraction and the relevant parameters for signal and background distributions.  The algorithm usually needs a numerical computation of the posterior, which we work out explicitly in a benchmark scenario of a simplified $pp\to hh \to b\bar b \gamma \gamma$ search.   We \change{perform a posterior predictive check on the results and we} show how the signal fraction is correctly extracted from the sample, as well as many parameters in the signal and background distributions.  The presented framework could be used for other searches such as $pp\to ZZ,\ WW,\ ZW$ and pair of Leptoquarks, among many others. 

\end{abstract}

\maketitle

\section{Introduction}
LHC is a challenging machine.  It has discovered the Higgs Boson \cite{higgs_disc1, higgs_disc2}, it has studied in detail the top \cite{atlastop,cmstop} and the Electro-Weak bosons \cite{atlasew1, cmsew1, cmsew2}, it is expected to be sensitive to di-Higgs production \cite{prospects,prospects2} --which probably would be its last test for the Standard Model (SM)--, it has pushed the limits and discarded many New Physics (NP) theories \cite{exotic}, and many other achievements. Nowadays, since a few years ago, it is mainly dedicated to scrutinizing in detail the observables and looking for NP in all of them and beyond with so far no positive results.  Despite all of this, the LHC is one of the main machines for looking and discovering NP, and the challenge is to find and develop the best ways to pursue this goal.

One of the directions to increase the sensitivity for new findings in the data analysis is using Machine Learning (ML) techniques.  Since the last decade, with the advent of the new ML algorithms and the new available hardware, these techniques have revised existing observables and created new ones that could explore in a previously unthinkable detail the data.  The ML industry has developed hundreds of algorithms, \change{ most of them corresponding to supervised and unsupervised learning.   At the LHC, however, the labelling of the data is unknown and the main way to utilize the supervised algorithms} is by training them through modeling of the data using Montecarlo (MC) simulations. This does not represent any problem, unless the simulation has some bias.  It is very difficult to tell whether and to what extent this may be the case, but nonetheless it is interesting to explore techniques in which one can reduce the impact of the MC simulations.  One possible path in this direction is to use {\it unsupervised} ML to learn from the data.

In this work we explore an unsupervised ML technique within a Bayesian inference framework. We study a general scenario of a signal consisting of decaying particles in the final state and its corresponding background, which is a situation that occurs often at the LHC.  Observe that the decay products in the signal would have different distributions to that of their corresponding particles at the background.  Therefore, the key feature is to observe that being more than one decaying particle yields a relationship between these decays at the event-by-event level that codifies relevant information about the physical system.  As for instance, the signal fraction in the sample and the preferred values for the parameters in signal and background distributions.  In the ML industry this scenario is called a mixture model and it has been extensively studied in the literature (see \cite{bishop} and references therein), whereas it has also been applied in LHC physics \cite{metodiev,lamagna,4tops,spannowsky,manu1}. \change{Searches by CMS in Refs.~\cite{cms1,cms2} perform an analysis with many similarities to a mixture model, although not mounted on the same mathematical framework.}   A suitable tool to extract all the relevant information in the sample, taking profit of the knowledge codified at each event, consists in considering a generative model for the data and then inferring its latent variables using Bayesian inference.  This is the approach we propose in the present manuscript.  

This article is organized as follows.  In the next Section we present the problem, its phenomenological aspects and a more mathematical formulation.  We define the variables that are used, and we discuss which numerical methods could be used to tackle the Bayesian inference.  In Section \ref{benchmark} we apply the method to a simplification of an important search at the LHC: $pp\to hh \to b \bar b \gamma\gamma$.  We show explicitly how the algorithm can be deployed and used, and we study its outcome in inferring important aspects of the physical problem.  In particular, we show how the method can learn many aspects from the data and obtain a good estimation for the signal fraction in the sample.  We follow with a discussion Section in which we analyse many aspects relevant to the previously obtained results and the method in general.  We end with the conclusions in Section \ref{conclusions}.

\section{Statement of the problem}

\subsection{Physical and phenomenological aspects of the problem} 

Given a sample of events that correspond to a mixture of a signal consisting of two or more decaying particles and its background, we wish to extract from the data: {\it i)} the signal fraction in the sample; and {\it ii)} the relevant distributions for signal and background.  For the sake of clarity and concreteness let us consider a benchmark case of a signal with 2 decaying particles, say $A\to a_1 a_2$ and $B \to b_1 b_2$.  In any case, all conclusions are valid as well for the case of more decaying particles and/or more decay products.  Moreover, the more decaying particles are in the signal process, the better the method should work.  The background, on the other hand, would consist of the non-resonant final state $a_1 a_2 b_1 b_2$.  Since signal and background correspond to different processes, then their distributions on any variable would in general be different.  In particular, the distribution for the invariant masses $m_{a_1 a_2}$ and $m_{b_1 b_2}$ would be resonant for signal and non-resonant for the background, which would usually behave as exponentially decaying in the invariant mass.  Since each event is either signal or background, a probabilistic generative model to reproduce the data-set would be what is known as a mixture model.

We consider a sample of $N$ events and we want to extract information about the mixture fraction and the signal and background features.   In order to discriminate between signal and background, one needs their distributions to be different in such a way that their weighted sum unambiguously determines which is the fraction of each one in the sample and also the relevant parameters from their distributions.  The ideal case corresponds to infinite data and without the mentioned ambiguity, in which case it is straightforward to obtain the sought features.  In real scenarios, however, one has access to finite data and fluctuations play a non-negligible role in hindering the signal-background discrimination.  It is therefore important to find observables that reduce the potential ambiguity in the signal and background distributions, even in the presence of fluctuations.

In the particular case that we study, there is a subtlety that is worth observing it with some detail.  Since the particles $A$ and $B$ are physically created and then decay, then the $m_{a_1 a_2}$ and $m_{b_1 b_2}$ distributions are practically independent in signal.  In background this is usually not the case, however, after the selection cuts that define the signal region, in general one can take these observables as approximately independent also for background.  That is $m_{a_1 a_2}$ and $m_{b_1 b_2}$ are approximately independent for signal and the same for background.  The key point is to observe that if we mix signal and background in a mixture model, then they acquire a dependence.  Naively speaking: given the value of $m_{a_1 a_2}$, the probability density function (PDF) for $m_{b_1 b_2}$ has a non-trivial dependence on $m_{a_1 a_2}$ and on  the parameters of the distributions, including the signal fraction.  Therefore the outcome of the measurements of $m_{a_1 a_2}$ and $m_{b_1 b_2}$ for {\it each} event provide relevant information on the sought parameters and thus contribute to reduce potential ambiguities between signal and background distributions.  This dependence, and the procedure to take full advantage of the information encoded in the data at an {\it event-by-event} level, is what we discuss and describe below within a Bayesian framework.

\subsection{Mathematical formulation within a Bayesian framework}
\label{mate}

\begin{figure}
    \centering
    \includegraphics[width=0.7\textwidth]{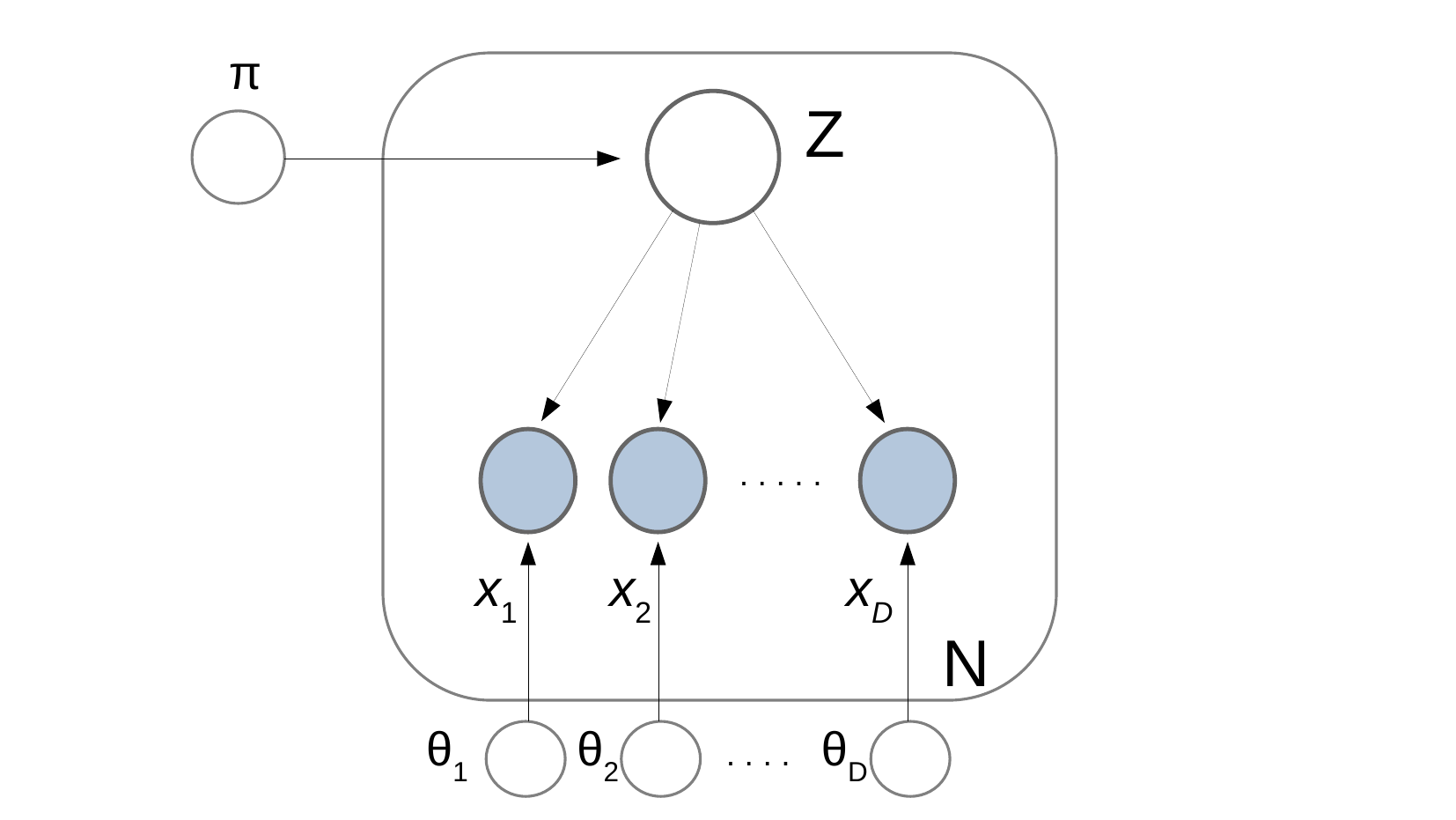}
    \caption{Graphical Model for the generic proposed model.  The latent variable $Z$ is sampled according to the parameter $\pi$ at each event to determine to which class belongs the event (signal or background in our benchmark scenario).  Then each one of the conditional independent random variables $x_i$ is sampled according to their $\theta_i$ parameter.  (These $x_i$ $(i=1,2)$ are the two invariant masses in our benchmark scenario.) This sampling procedure is repeated $N$ times.   Observe that in this model the values of the $x_i$ are the only observed ones, whereas all others are inferred from the generated data.  In the Graphical Model language, shaded and empty circles refer to observed and latent random variables, respectively.}
    \label{graphicalmodel}
\end{figure}

We consider that the data has been sampled from a mixture model and pursue the goal of determining the parameters of the model given the data.  \change{We use a Graphical Model as in Fig.~\ref{graphicalmodel} to represent the probabilistic model.  Graphical Models is a theoretical tool from Statistics, often applied in the ML industry, that is very useful to visualize and understand the structure and the dependence in a probabilistic model.  In this framework, each random variable is represented with a circle, and arrows indicate that the realization of one random variable affects another random variable.  Empty circles indicate that a random variable is not observed (latent random variable), and  filled circle that it is observed (observable).  A line surrounding many random variables indicate that these should be sampled many times, $N$ times in Fig.~\ref{graphicalmodel}. The idea is that using the observed random variables, one can infer the unobserved (latent) random variables.   Graphical Models, although at this level provide no more than a practical and useful visualization of a model\footnote{Something similar happens with Feynman diagrams in high energy physics:  They are a crucial tool to help visualize processes and to write down sophisticated calculations.}, result compelling in enhancing the usefulness of a Bayesian framework.  More details about its definitions and applicability can be found in Ref.~\cite{bishop}.
}

We assume that the data $X$ consists of $N$ events with $D$ observed random variables in each event, which we name $x^{(n)}_{1...D}$ for the $n$-th event.  Each event can be of any of the classes determined by sampling the latent (unobserved) categorical random variable $Z$.  In our benchmark case discussed above and studied in next section, we have 2 classes ($Z$ = signal or background) and $D=2$ observations ($m_{a_1 a_2}$ and $m_{b_1 b_2}$).  The parameters for the distribution of the random variables $Z$ and $x^{(n)}_i$, which are collectively called $\theta$, are sampled once for the whole sample.  For $Z$ these parameters could be sampled from a Dirichlet distribution (Beta distribution for the benchmark case), and for the $x^{(n)}_i$ depends on the specific feature represented by the random variable.  In particular, observe that the parameters that rule the distribution for $Z$ in the benchmark scenario, $\pi_S$ and $\pi_B=1-\pi_S$ indicate the expected fraction of signal and background in the sample, respectively.

Observe that in this model the random variables $x^{(n)}_{1...D}$ are conditionally independent given $Z$,
$$p(x^{(n)}_1,..., x^{(n)}_D| Z ) = p(x^{(n)}_1| Z ) \,...\, p(x^{(n)}_D| Z ) $$
but they are not independent if $Z$ is not observed.  Since $Z$ is not observed, their dependence arises from their class assignment, providing a leverage to enhance the inference over the parameters given the data $X$.

We have therefore the Bayesian problem of inferring the parameters $\theta$ given the data $X$,
\begin{equation}
    p(\theta|X) = \frac{p(X|\theta)\, p(\theta)}{p(X)}.
    \label{bayes}
\end{equation}
Here $p(\theta)$ is the prior, $p(X|\theta)$ is the likelihood for $\theta$, $p(X)$ is the evidence, and $p(\theta|X)$ is the posterior probability.  Finding out $p(\theta|X)$ is the objective of the problem, since this determines not only the probability density for the values of the signal fraction in the sample, but also it can tune the relevant parameters for the signal and background distributions using the data in the signal region.  This is a remarkable aspect of the method, since it can reduce the impact of Montecarlo tuning in the quest for physical quantities, as shown in next Section and discussed below.

\change{
It is worth noticing at this point that, although the likelihood $p(X|\theta)$ may be a well behaved and simple to express function, the posterior $p(\theta|X)$ is often sophisticated and also multimodal.  This multimodality may be intrinsic from the problem, but is also expected to have a contribution from the data $X$, which in principle has been sampled, is finite, and therefore it is not smooth.  This multimodality problem is well known, it has been largely studied (see for instance \cite{bw}) and, combined with the numerical difficulties of sampling from the posterior, and true parameters in a region of approximately degeneracy, yields that usually the maximums in the marginalized posterior do not match exactly all the true parameters of the data.  As a matter of fact, the mathematical formulation of Bayesian inference does not expect this, but instead that the posterior is compatible with the true parameters of the data and, if the workflow is successful, that the posterior evaluated at the true parameters is larger than the prior.  That is, that we have gained information throughout the process by using the data.
}

To solve Eq.~\ref{bayes} for the posterior we need the prior and the likelihood.  One key step to use this framework in order to reduce the Montecarlo impact on the final results, is to be able to write the likelihood using some robust knowledge that one has from the physical problem, and leaving some variables that would be adjusted by the data in the inferring process.  In the benchmark case studied in the article, we model that the signal comes from the convolution of a resonant Breit-Wigner with all the detector effects, which we approximate as a Normal distribution, whose parameters' exact values we leave as variables to be adjusted in the inferring process.  On the other hand, we also have the knowledge that the background can be approximated as exponential decaying with the invariant mass (any of them), since we know that the energy availability in the proton parton distribution functions is exponentially decaying.  The exact values of the corresponding parameters we also leave them as variables to be adjusted.  

Since in the model the $x^{(n)}_i$ are conditionally independent on $Z$, we describe their distribution as a mixture of direct products with the corresponding signal and background distributions.  For the case of only two classes ($k=2$, signal and background):
\begin{eqnarray}
\ln p(X|\theta) &=& \sum_{n=1}^N \ln \left( \pi_S \, p_S ({\bf x}^{(n)}|\theta_S ) +  (1-\pi_S) \, p_B ({\bf x}^{(n)}|\theta_B )    \right) \label{likelihood} \\
p_S({\bf x}^{(n)}|\theta_S ) &=& {\cal N}(x_1^{(n)}; \mu_1, \sigma_1) \times ...\times {\cal N}(x_D^{(n)}; \mu_D, \sigma_D)  \label{ps} \\
p_B({\bf x}^{(n)}|\theta_B ) &=& \mbox{Exp}(x_1^{(n)}; \lambda_1) \times ...\times \mbox{Exp}(x_D^{(n)}; \lambda_D) \label{pb}
\end{eqnarray}
where for the sake of concreteness we stuck to the idea that the signal and background are Normal- and Exponential-distributed, respectively.  Of course that this is valid for any other distributions.  Observe that in Eq.~\ref{likelihood} we have marginalized over the latent variable $Z$ since it is not observed (see Ch.9 in Ref.~\cite{bishop} for details).  Also observe that the parameters for the signal and background distributions are $\theta_S = \mu_1,\,\sigma_1,\, ...,\mu_D,\,\sigma_D$ and $\theta_B = \lambda_1,\,....,\lambda_D$, respectively. The other parameter belonging to the to-be-inferred parameters in $\theta$ is the signal fraction $\pi_S$.   The maximization of the likelihood in Eq.~\ref{likelihood} is in general an intractable task because, although all distributions belong to the exponential family, the sum inside the logarithms precludes obtaining a closed form solution for the maximum likelihood estimation.  This is a well known problem for mixture models in the Machine Learning field, but there are many workarounds, some of which we discuss below.

\subsection{Numerical methods}
\label{numerical} 

The maximization of the likelihood in a mixture model is a long time problem in the Machine Learning industry.  

One very efficient and classic method is the Expectation-Maximization or EM algorithm \cite{EM, bishop}.  This algorithm consists in taking the expectation (E) of the log-likelihood with respect to $Z$ and then maximizing (M) with respect to the $\theta$ parameters in an expression that results not exact, but tractable.  If this procedure is repeated a few times, the likelihood always increases and finds a local maximum.  The drawback of this algorithm is not only that it could not be finding the global maximum, but more important that it does not provide a posterior probability density function.  And since fluctuations play an important role in the problem we are dealing with, we are interested in reconstructing a posterior for the $\theta$ parameters.  In any case, if using this algorithm, one should be cautious because usually the region of analysis (signal region) will be bounded in the invariant masses, and then the Normal and Exponential distributions should be replaced by the Truncated Normal and Truncated Exponential, respectively, whose normalization factors play a crucial role in maximizing the likelihood.  We have verified that despite this potential complication, the EM algorithm can be applied to the problem.  

Another method for obtaining samples from the posterior is dynamic nested sampling.  Nested sampling \cite{nested} is a method whose original purpose is to estimate the evidence by integrating iso-likelihood contours.  It works by computing the volume ${\cal X}$ in the prior space that corresponds to a given threshold in likelihood, and then integrates the likelihood iso-contours over $d{\cal X}$.  Dynamic nested sampling is an adaptation to dynamically modify the number of live points in the large weight regions in order to have a better reconstruction of the posterior.  The algorithm can also be used to efficiently estimate the posterior by assigning the corresponding importance weight to each sample in the evidence computation.  In the next Section we use {\tt Dynesty} \cite{dynesty}, a Python library that implements these techniques, to run an inference process for the posterior probability on the parameters of the model given the data.

\change{

\subsection{Posterior predictive check}
\label{posterior_predictive}

In Bayesian Inference, once inference has been performed on a given data and with a given model, there are a variety of safeguards against potential faults in the procedure, such as for instance mismodelling, misspecifications in the model, model misfit, and/or numerical bugs among others.  There is not a systematic way to select which checks should be performed on a model, but running a few sanity checks is a good protection versus possible mistakes and biases in the results \cite{bw,gelman}.  There is also no standard way to determine that if a given check fails, then the model requires adjustments.  There is an interplay between costs and benefits that, depending on whether the model captures or not the relevant aspects encoded in the data, may yield to invest in improving the model.   A detailed discussion on these aspects can be found in Ref.\cite{bw}.

If the inference is numerical, like in a Markov Chain MonteCarlo or nested sampling \cite{nested}, then the corresponding convergence tests shall be performed.  It is also important to run a toy model in which all distributions are well known and controlled, and verify that the procedure works as expected.   Finally, a general check in Bayesian inference is the {\it posterior predictive check}, which we discuss below.  Refs.~\cite{bw} and \cite{gelman}, and references therein, have a detailed discussion on different sanity and validations checks on Bayesian workflow.

The posterior predictive check consists in using the results of the inference to test whether the prediction for the data that is obtained from the posterior is compatible with the true data.  That is, the posterior provides a PDF for the parameters of a probabilistic model for the data, therefore one can generate a replicate of the data with the probabilistic model by sampling the model parameters from the posterior, and then a replicate of the data from the model.  Once this is done, one can compare whether the replicated data is statistically compatible to the true data.  Needless to say that there is no standard way of doing this: comparing a finite sample of true data to a replicate sampled from a model whose parameters are sampled from a given PDF it is a task that does not have a unique answer.  We choose to use in this work a posterior predictive check based on Ref.~\cite{blei-causal} whose details are described in the next pargraphs.

Given a data set of observations $X$, a part of it is held-out ($X_{held}$) in the inference process:
$$ X = X_{obs} + X_{held} .$$
Let the posterior over $\theta$ inferred with $X_{obs}$ be $p(\theta|X_{obs})$.  From here we can generate a replicate data set ($X_{rep}$) whose similarity to the true data set will depend on the procedure and model used in the inference: The better is the model and the inference procedure, the more alike should be the replicate data set to the true data.  The probability distribution for the replicate data would be \cite{blei-causal}
\begin{equation}
    p(X_{rep} | X_{obs}) = \sum_Z \int p(X_{rep}, Z|\theta) \, p(\theta | X_{obs} )\, d\theta .
    \label{replicates}
\end{equation}
Observe that in this expression we are integrating out the posterior and the latent variables.
This is a PDF that we can obtain using the samples of the posterior obtained in solving the inference problem.   Since we have the actual value for $X_{held}$, we can compute its probability $p(X_{held} | X_{obs})$.   We define the posterior predictive score as the probability that replicate data has less probability that the held-out data,
\begin{equation}
    \mbox{predictive score} = p\biggl(  p(X_{rep}|X_{obs}) < p(X_{held}|X_{obs}) \biggr) .
    \label{predictive_score}
\end{equation}
Or in words, the area under the replicate PDF in Eq.~\ref{replicates} where the PDF value is below the value taken at $X_{held}$.   An ideal predictive score is $0.5$.  Above $0.5$ means that the model has some misspecifications, that predicts most of the data where actually $X_{held}$ is, but also in broader regions where there is no held-out data.  A predictive score below $0.5$ means that the model predictions are biased with respect to $X_{held}$.  Depending on how difficult it is expected to find a model that reproduces the data, one may use different specific (subjective) thresholds to consider a model suitable.   We expand this discussion in Sect.~\ref{posterior_predictive2} while applying it to a real example.
}

\section{A simplified benchmark scenario for \texorpdfstring{$pp\to hh\to b\bar b \gamma\gamma$}{pp to hh to b b gamma gamma} }
\label{benchmark}

\begin{figure}[th]
    \centering
    \begin{minipage}[c]{\textwidth}
    \centering
    \includegraphics[width=0.45\textwidth]{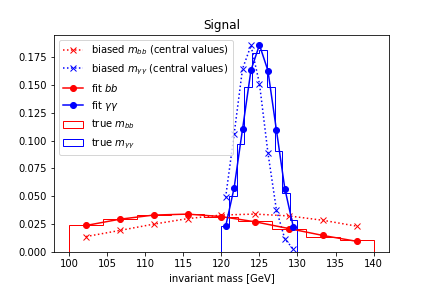}
    \includegraphics[width=0.45\textwidth]{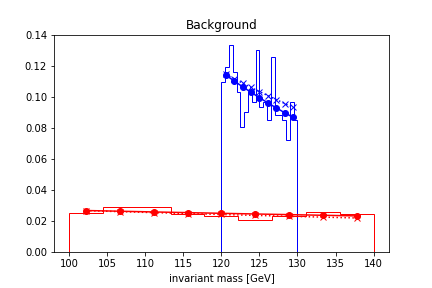}
    \end{minipage}
    \caption{Solid: Fit to Normal (signal) and Exponential (background) distributions of the $m_{b\bar b}$ and $m_{\gamma\gamma}$ data, as coming from the {\tt Madgraph} + {\tt Pythia} + {\tt Delphes} simulations.  Dashed: Normal and Exponential distributions with their corresponding parameters shifted as a method to emulate a Montecarlo that does not exactly match the data.  These shifts corresponds to the central values of the biased priors for the $\mu_{b\bar b},\ \sigma_{b\bar b},\ \mu_{\gamma\gamma},\  \sigma_{\gamma\gamma}$, $\lambda_b$ and $\lambda_{\gamma}$ parameters.}
    \label{sandb}
\end{figure}

To analyse the power of the proposed method we test it in a simplification of a realistic scenario of the LHC di-Higgs search in the $b\bar b + \gamma\gamma$ channel.  It is not the intention of this analysis to do a real study on this search, but to show the potential of the algorithm in a simplified scenario of a real and relevant physical case.  Actually we only consider gluon fusion production and the irreducible non-resonant background $pp\to b\bar b \gamma\gamma$, neglecting vector boson fusion production and backgrounds such as $t\bar t h$, $hb\bar b$ and $\gamma\gamma jj$ among others \cite{atlas, cms}.

\change{\subsection{Bayesian inference}}
\label{bi}

We have concatenated Madgraph \cite{madgraph} + Pythia \cite{pythia} + Delphes \cite{delphes} to simulate matrix-level, radiation, showering, hadronization, and detector level for signal and background.  In signal we have simulated $pp \to hh \to b\bar b \gamma\gamma$, whereas in background $pp \to b\bar b \gamma\gamma$.  We have collected all data and the corresponding scripts for this Section results in a {\tt GitHub} repository \cite{github}.  We have simulated enough signal and background to have more than 1k events of each one that fulfill at detector level the reconstructed $b\bar b \gamma\gamma$ final state with $100\mbox{ GeV} \leq m_{b\bar b} \leq 140 \mbox{ GeV}$ and $120\mbox{ GeV} \leq m_{\gamma\gamma} \leq 130 \mbox{ GeV}$, as guided by the signal regions in Refs.~\cite{atlas,cms}.  Observe that the width in energy for the signal region in $m_{b\bar b}$ and $m_{\gamma\gamma}$ is quite different because of the different uncertainties in reconstructing bottoms and photons at the detector level.  

\begin{figure}[th!]
    \centering
    \includegraphics[height=0.8\textheight]{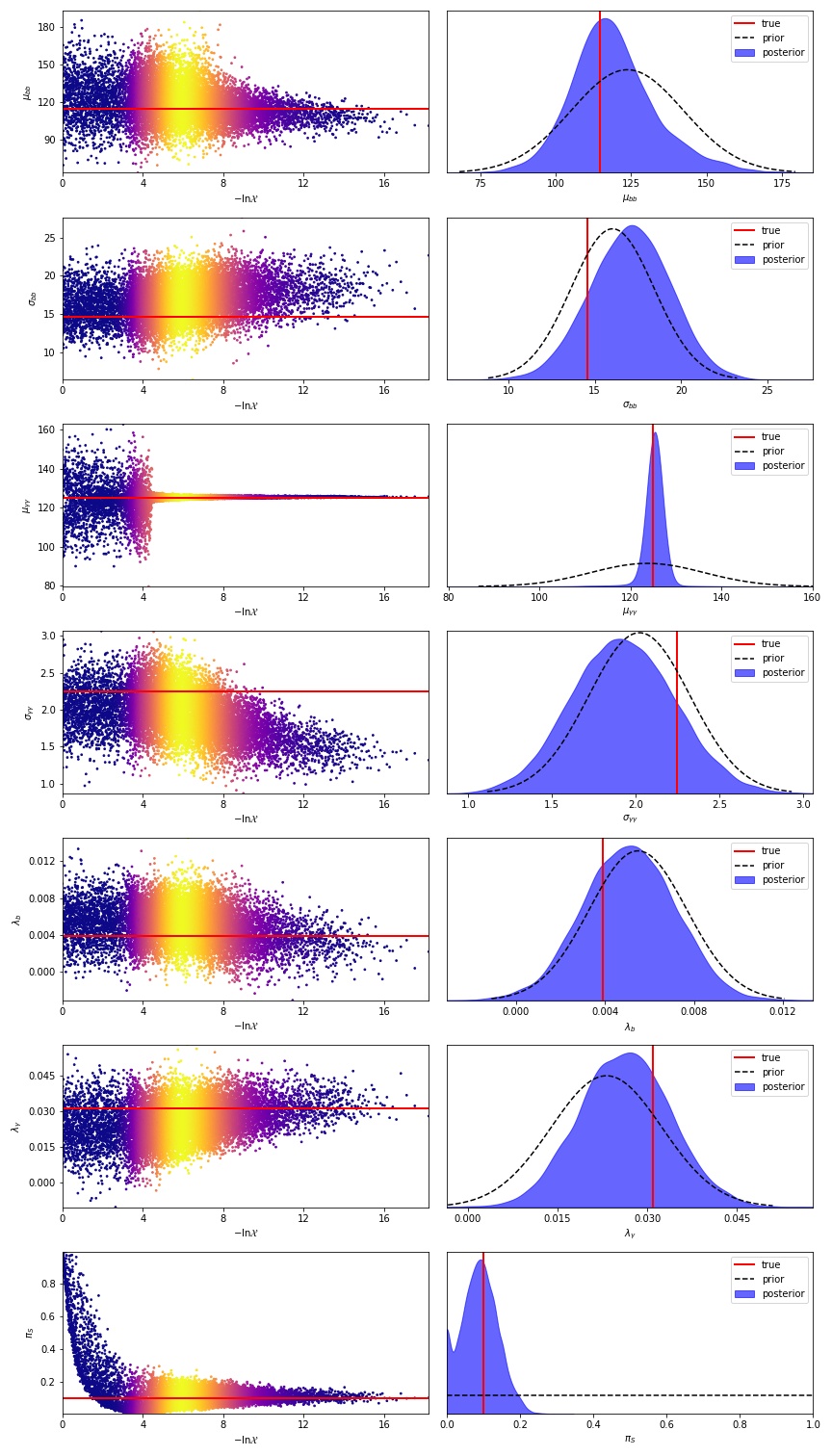}
    \caption{Dynamic nested sampling through {\tt Dynesty} to infere the values of the 7 parameters of the model ($\mu_{b\bar b},\ \sigma_{b\bar b},\ \mu_{\gamma\gamma},\ \sigma_{\gamma\gamma},\ \lambda_b,\ \lambda_\gamma$ and $\pi_S$) for a data set with 10\% signal.  Left column shows the sampling evolution as the likelihood increases and the prior volume (${\cal X}$) shrinks.  The brighter the color of each point, the larger is its corresponding weight.  Right column shows the posterior distribution of each parameter after the dynamic nested sampling inference process.  We also show in dashed the prior distribution of each parameter and in red its true value in the fit of the labeled data.}
    \label{fulldynesty}
\end{figure}    

The algorithm we are presenting has the ability to learn from the data the signal fraction in the sample, as well as to learn the distributions in each one of these classes.  To test this feature, we emulate the scenario in which the Montecarlo predictions are slightly biased from the data.  In order to do this we have fitted the labeled data to Normal (signal) and Exponential (background) distributions, and then proposed a prior distribution for the corresponding distribution parameters that is shifted from their fitted values.   The graphical outcome of this procedure is depicted in Fig.~\ref{sandb}.  Observe that we are plotting the distributions that correspond to the central values of the biased prior distributions.  In Fig.~\ref{fulldynesty} (explained below) one can see the prior distribution of each parameter.
\begin{figure}[th!]
    \centering
    \begin{minipage}[c]{\textwidth}
    \centering
    \includegraphics[width=0.31\textwidth]{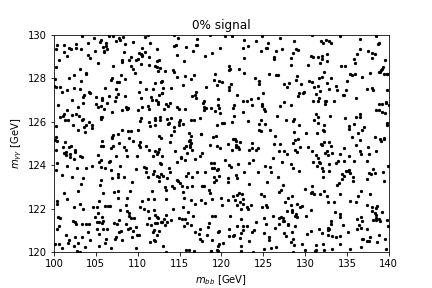}
    \includegraphics[width=0.31\textwidth]{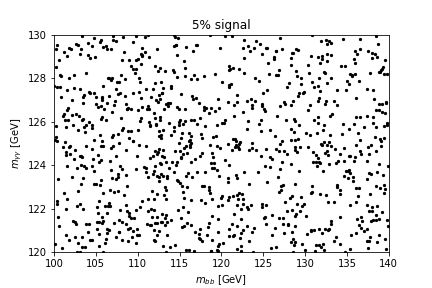}
    \includegraphics[width=0.31\textwidth]{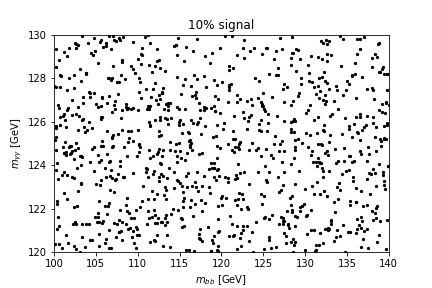}
    \end{minipage}
    \begin{minipage}[c]{\textwidth}
    \centering    
    \includegraphics[width=0.31\textwidth]{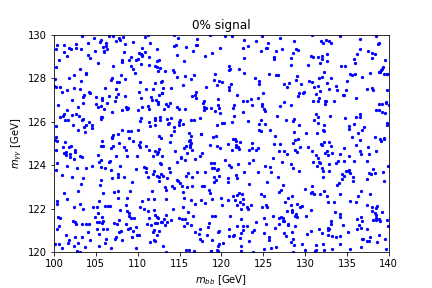}
    \includegraphics[width=0.31\textwidth]{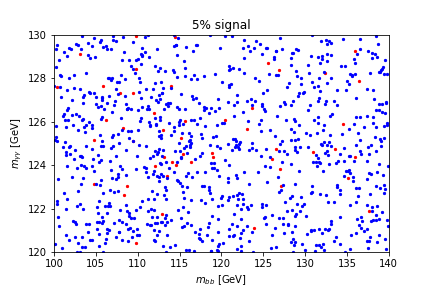}
    \includegraphics[width=0.31\textwidth]{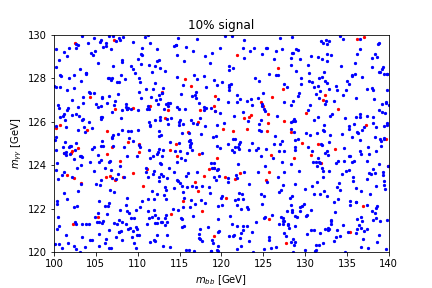}
    \end{minipage}
    \begin{minipage}[c]{\textwidth}
    \centering    
    \includegraphics[width=0.31\textwidth]{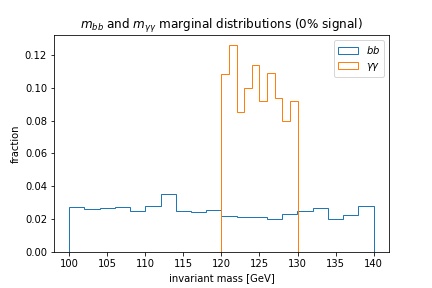}
    \includegraphics[width=0.31\textwidth]{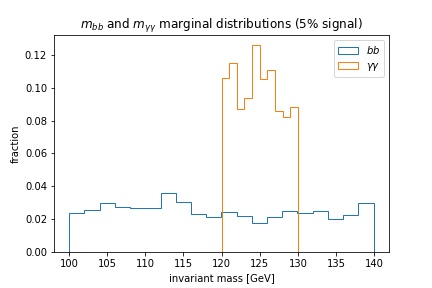}
    \includegraphics[width=0.31\textwidth]{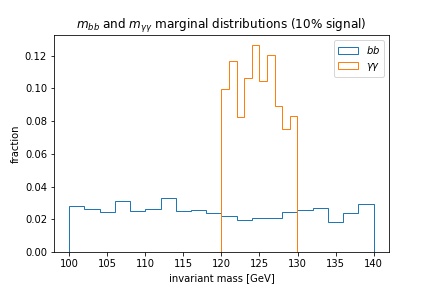}
    \end{minipage}
    \begin{minipage}[c]{\textwidth}
    \centering    
    \includegraphics[width=0.31\textwidth]{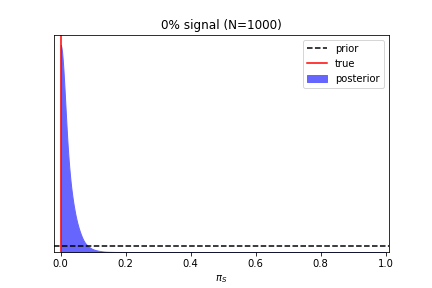}
    \includegraphics[width=0.31\textwidth]{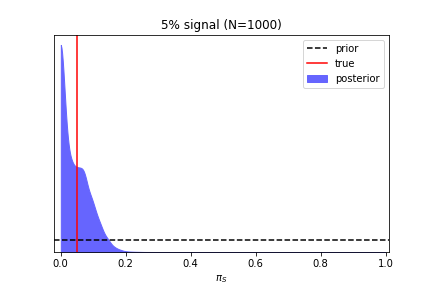}
    \includegraphics[width=0.31\textwidth]{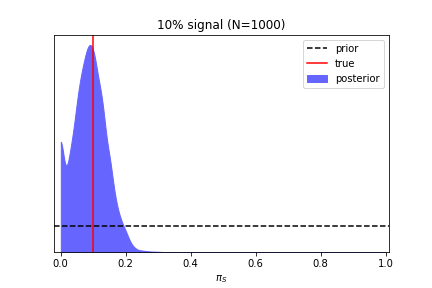}
    \end{minipage}
    \caption{Bayesian extraction of signal fraction in a LHC simplified two-components simulated sample of $N=1000$ events of $b\bar b \gamma\gamma$: $pp\to hh \to b\bar b \gamma \gamma$ signal, and non-resonant $pp\to b\bar b \gamma \gamma$ background.  Left, middle and right columns correspond to a signal fraction of 0\%, 5\% and 10\%, respectively. Upper row corresponds to the data as it would be seen at the LHC. The second row unveils which events correspond to signal (red) and which to background (blue).  The third row indicates the invariant mass distribution for $b \bar b$ and for $\gamma\gamma$ marginalizing on the other distribution.  Bottom row shows the result of the inference process for $\pi_S$, which takes advantage of the $m_{b\bar b}$-$m_{\gamma\gamma}$ correlation in each event through the presented Bayesian algorithm.}
    \label{1000}
\end{figure}

\begin{figure}[ht]
    \centering
    \begin{minipage}[c]{\textwidth}
    \centering
    \includegraphics[width=0.31\textwidth]{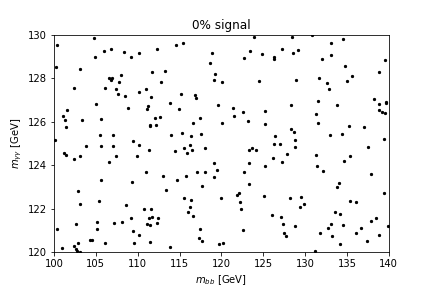}
    \includegraphics[width=0.31\textwidth]{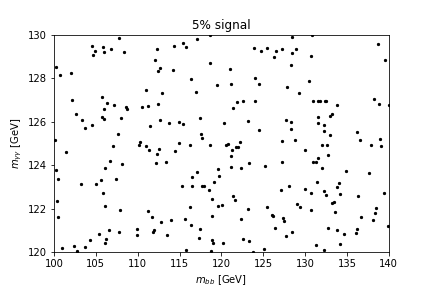}
    \includegraphics[width=0.31\textwidth]{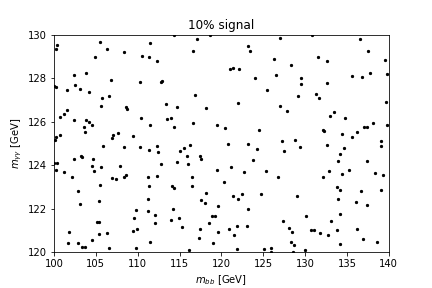}
    \end{minipage}
    \begin{minipage}[c]{\textwidth}
    \centering    
    \includegraphics[width=0.31\textwidth]{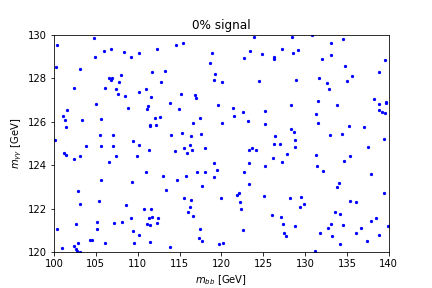}
    \includegraphics[width=0.31\textwidth]{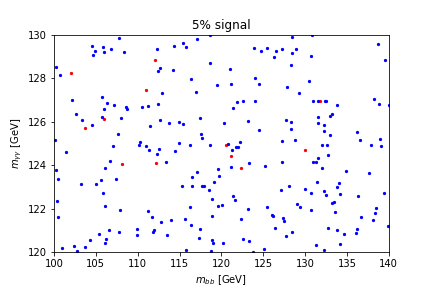}
    \includegraphics[width=0.31\textwidth]{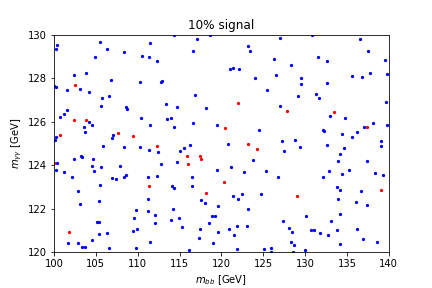}
    \end{minipage}
    \begin{minipage}[c]{\textwidth}
    \centering    
    \includegraphics[width=0.31\textwidth]{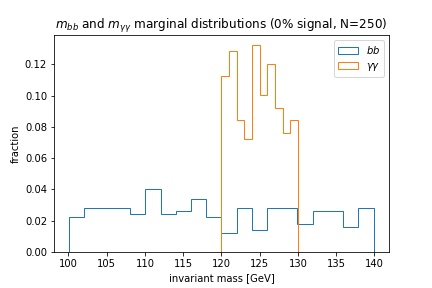}
    \includegraphics[width=0.31\textwidth]{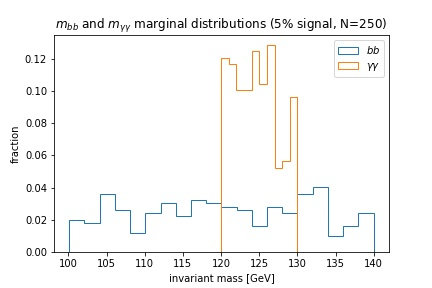}
    \includegraphics[width=0.31\textwidth]{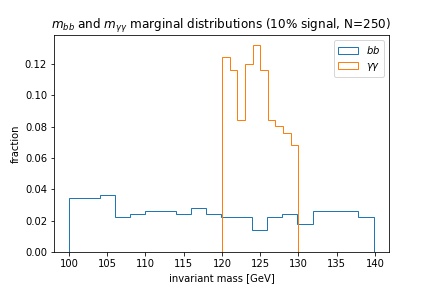}
    \end{minipage}
    \begin{minipage}[c]{\textwidth}
    \centering    
    \includegraphics[width=0.31\textwidth]{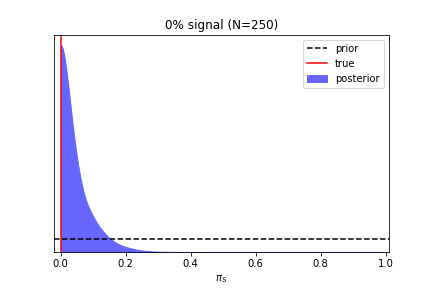}
    \includegraphics[width=0.31\textwidth]{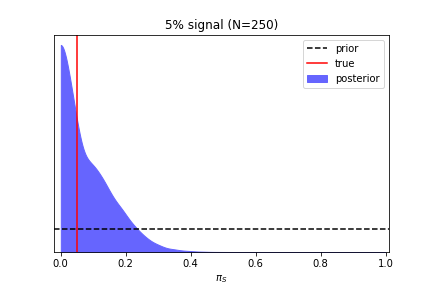}
    \includegraphics[width=0.31\textwidth]{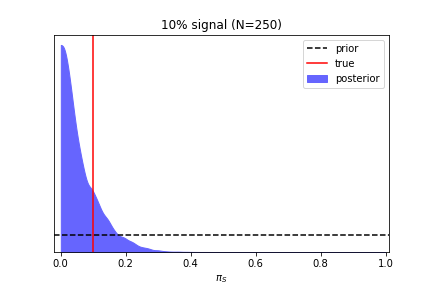}
    \end{minipage}
    \caption{Idem as Fig.~\ref{1000}, but for $N=250$ total events in the sample.}
    \label{250}
\end{figure}

We have run Bayesian inference processes for the unknown parameters of the model, namely $\mu_{b\bar b},\ \sigma_{b\bar b},\ \mu_{\gamma\gamma},\ \sigma_{\gamma\gamma},\ \lambda_b,\ \lambda_\gamma$ and $\pi_S$, using dynamic nested sampling through the library {\tt Dynesty}.  We have used the likelihood as defined in Eq.~\ref{likelihood}-\ref{pb}, but using Truncated Normal and Truncated Exponential distributions in the signal region defined in the $m_{b\bar b}$-$m_{\gamma\gamma}$ plane for the values previously mentioned.   We have run the algorithm for samples of $N=1000$ and $N=250$ total events with signal's fraction of 0\%, 5\% and 10\%.  The outcome of one of these runs ($N=1000$ events with 10\% signal, i.e.~$\pi_S=0.1$) is shown in Fig.~\ref{fulldynesty}.  In the left panel it can be seen the evolution of the sampling process as a function of the prior volume ${\cal X}$, and in the right panel the outcome of the posterior for each one of the inferred parameters.  The prior distribution for each parameter is shown in a black-dashed line, and the Normal means are allowed to explore values beyond the selected signal region.  Their central values are biased from the true-level, as explained above.  As it can be seen, the means $\mu$'s are learned very efficiently from the data, slightly less the $\lambda$'s, and the $\sigma$'s are learned slightly in the wrong direction, 
\change{but compatible with the true values.  This is in concordance with the discussion in Sect.~\ref{mate} concerning the multimodality and parameters degeneracy in the posterior distribution.  Moreover, one can get an understanding for the reasons behind this behavior in this case.  In the problem we are dealing with, $m_{b\bar b}$ is allowed to span from 100 to 140 GeV and its signal width is as large as $\sigma_{b\bar b} \sim 15$ GeV, whereas its background decaying rate is as small as $\lambda_b \sim {\cal O}(10^{-3})$ GeV$^{-1}$.   Therefore, we can recognize a partial degeneracy as follows: one could have a slight increase in $\sigma_{b\bar b}$ at the price of making steeper the decay in its background (increasing  $\lambda_b$) and slightly correcting the signal fraction.  (And vice versa).  For the case of $m_{\gamma\gamma}$ the signal width is slightly smaller in comparison to its allowed span range (2 GeV in the range 120-130 GeV), however, being the signal fraction as small as 10\%, this behavior is also present, but less pronounced.  Observe that this pattern is observed in both $m_{b\bar b}$ vs.~$\lambda_b$ and $m_{\gamma\gamma}$ vs.~$\lambda_\gamma$ in Fig.~\ref{fulldynesty} when comparing the posterior to the true values. Note that it is not by chance that the data lie in such a critical range of their values, but precisely because this is the signal region in the di-Higgs searches.  The previous discussion can also be visualized using the upper rows in Figs.~\ref{1000} and \ref{250}.

To verify that the above features are not a matter of statistics, we have generated three times the data than in Fig.~\ref{fulldynesty}, and found that still with more data there is not a significant improvement in the results. The results for this run can be found in \cite{github}.  Whereas using a toy model to generate fake data with new true parameters that break the above degeneracy, produces a better convergence with just half of the events (see Appendix).
} 

What is compellingly learned in the inference that yields Fig.~\ref{fulldynesty} is the expected signal fraction in the sample (bottom panel): starting from a totally unknown fraction (uniform distribution) the algorithm learns a posterior probability for $\pi_S$ whose mass is accumulated around the true value.  This results is obtained for other signal fractions, and with less events as well, and is one of the main results in this work.  \change{ A few remarks shall be mentioned on this result.  Although to enhance the power of the method we have left the fraction prior as flat, in a real scenario one would probably use prior information about the signal fraction from the Montecarlo and, therefore, less parameter space would be discarded, but also the posterior would have less spread.  It is also true that in our understanding the prior knowledge on the other parameters that are not the signal fraction have different origin and we consider them physically more robust.  This is because they use information such as the Normal and Exponential nature of the signal and background distributions, the measured decaying rate shape of the background invariant masses, the Higgs mass, the measured Jet Energy Scale uncertainties --which give an idea of what to expect for $\sigma_{b\bar b}$--, and the resolution of the electromagnetic calorimeter, among others.

}

In order to have a better visualization of the scope of the algorithm we show in Fig.~\ref{1000} ($N=1000$) and Fig.~\ref{250} ($N=250$) a summary of the results for signal fraction 0\%, 5\% and 10\% (from left to right columns).  We show in the upper plots the distribution of events in the $m_{b\bar b}$-$m_{\gamma\gamma}$ plane without distinguishing signal from background to gain an insight of how difficult it would be to recognize the signal fraction and other features.  In the second row we show which events correspond to signal.   The third row shows the $m_{b\bar b}$ and $m_{\gamma\gamma}$ distributions by marginalizing in the other one.  That is, these plots show how the data would look if it is not taken at the event-by-event level, as needed by the algorithm.   The bottom row in the figure shows the outcome of the inference process for the expected signal parameter $\pi_S$ and the corresponding true signal fraction.  As it can be seen, the level of agreement of the posterior distribution mass with the true value is very good. Also important is to observe that the $\pi_S$ posterior distribution can discard an important range of values that were previously allowed with the prior.

It is worth stressing at this level not only the blurred and biased information that we have poured into the data to let the algorithm extract so many features, but also the relatively few events and how difficult is to observe any pattern in the data (upper row in Figs.~\ref{1000} and \ref{250}).  As discussed above, one of the main strengths of the algorithm lies in looking at the data event-by-event and gradually gathering the correct information to build up the posterior. \change{ Moreover, as support of the above we cite the analyses by CMS \cite{cms1,cms2} which consist in a 3-dimensional fit in three invariant masses (equivalent to take all the information at the event-by-event level) and show a 20-30\% improvement with respect to the same analysis in which events are first selected by fitting two dimensions, and then those selected events are analyzed using the other dimension. }

\change{
\subsection{Posterior predictive check}
\label{posterior_predictive2}

In the following paragraphs we discuss some of the checks that should be done when solving an inference problem, applied to the previously simplified example.  

The first thing that should be noticed is that before embarking into the simulated data used in the previous section, we have tested the whole procedure within a controlled environment with fake data, usually known as a toy model.  We have generated fake data with specific Normal and Exponential distributions for each of the classes and we have correctly extracted the corresponding parameters of the PDF with which the data was generated.  We summarize the results in the Appendix \ref{toy} and present all the scripts and numerical data in \cite{github}.

On the other hand, for the mentioned controlled environment as well as for the simulated data from the previous paragraphs, when solving an inference problem using a dynamic nested sampling method such as with {\tt Dynesty}, one has to cross-check that convergence criteria for the numerical methods are correctly satisfied.  We have run {\tt Dynesty} with the following command and options
\begin{eqnarray}
\mbox{{\small {\tt DynamicNestedSampler.run\_nested(dlogz\_init=0.1, nlive\_init=500, maxiter\_init=5000, nlive\_batch=50,}}} \nonumber \\
\mbox{{\small {\tt n\_effective=10000, maxbatch=50),} }} \nonumber
\end{eqnarray}

\noindent where {\tt DynamicNestedSampler} is a {\tt Dynesty} object \cite{dynesty}.   This command, in addition to the set up of the run, indicates to stop when the effective number of samples in the posterior ({\tt n\_effective}) reaches 10000, since in this problem we pursue to construct a good posterior.   More details on the other set up options may be found in Ref.~\cite{dynesty}.   In all runs we have verified a correct convergence of the sampling methods, which is a sanity check for the posterior extracted with this numerical method.  More details and relevant results and plots can be found in Ref.~\cite{github}.

In addition to the above mentioned basic validation checks for the inference procedure we have performed a posterior predictive check, as described in Section \ref{posterior_predictive}.  We show the results for the case of 10\% signal in 1000 events.  We have used other 200 events as a held-out sample ($X_{held}$) that is not used in the inference.  This is equivalent to a sample of 1200 events in which 200 events are randomly held-out and not used in the inference process.

The procedure goes as follows.  Once the inference problem is solved using the proposed probabilistic model for the data, we have a posterior for the parameters in the model.  This posterior is the one shown in Fig.~\ref{fulldynesty} and depends on the data, the probabilistic model, and the inference procedure.  With this posterior we can sample parameters for the probabilistic model and then sample replicate data points ($m_{b\bar b}$ and $m_{\gamma\gamma}$ pairs) using Eq.~\ref{replicates}.  Observe that we can generate replicate data points at will, and that the compatibility of their distribution with the true data set distribution depends on the whole inference procedure.  This is what is studied in the posterior predictive check.

In our case, we have generated one million ($m_{b\bar b}$, $m_{\gamma\gamma}$) replicate pairs and divided the $m_{b\bar b}$--$m_{\gamma\gamma}$ plane into a 20x20 grid.  We have binned this replicate data set into the grid and obtained a numerical PDF for the replicate data set.  Therefore, using Eq.~\ref{predictive_score} and the value for each data point in $X_{held}$ we can compute the predictive score for the current problem, which yields 0.49.  This indicates that the there is a very good compatibility between the replicate data set and the true data set.  We show in Fig.~\ref{posteriorpredictiva} the replicate data set PDF and the $X_{held}$ data points in the same plot in the $m_{b\bar b}$--$m_{\gamma\gamma}$ plane, where the agreement can also be visually assessed.  It is worth noting, in any case, that the replicate data set PDF, as well as the true data distribution have little signal fraction and are mostly driven by the background, henceforth they do not have prominent shapes, but rather a smoothly changing density in the $m_{b\bar b}$--$m_{\gamma\gamma}$ plane.

\begin{figure}[th]
    \centering
    \begin{minipage}[c]{\textwidth}
    \centering
    \includegraphics[width=0.65\textwidth]{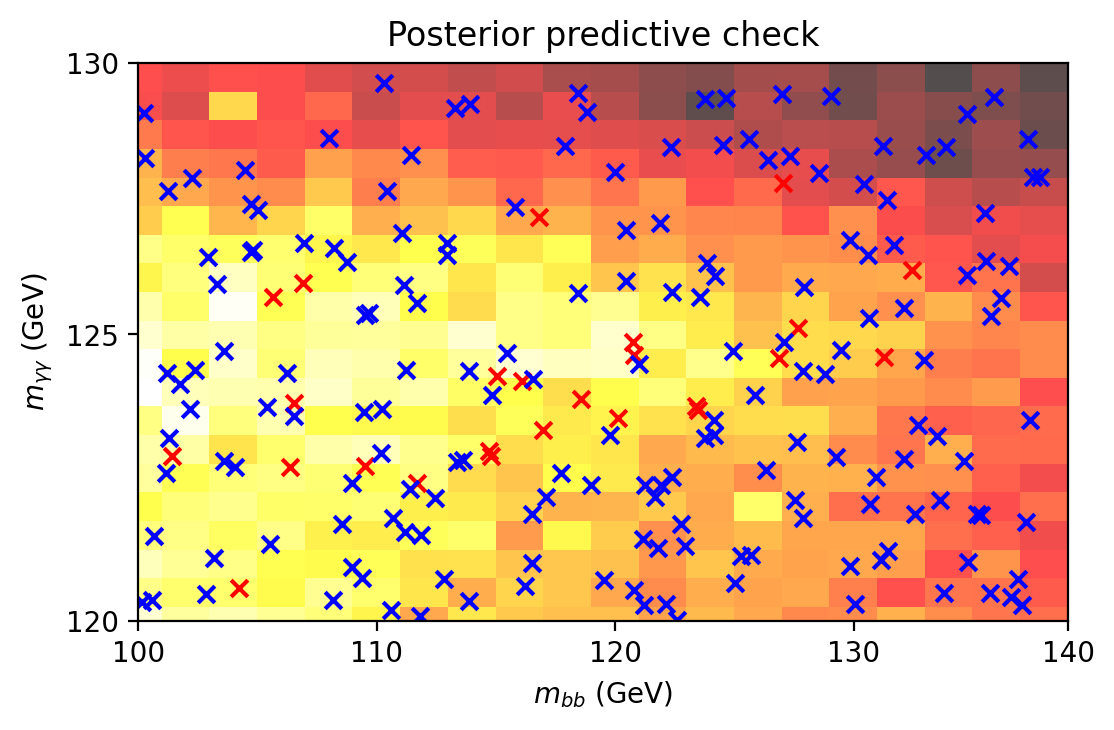}
    \end{minipage}
    \caption{\change{Posterior predictive check.  The background 20x20 coloured grid represents the PDF of a replicate of the data generated with the inferred posterior for the parameters of the model, as explained in Sect.~\ref{posterior_predictive}.  The crosses represent 200 held-out events (just for curiosity we have labeled blue for background and red for signal, but this information is not used).  The background PDF has been extracted without seeing the held-out events, and testing whether these held-out events are compatible with the replicated data PDF is a test for the whole Bayesian workflow.  In this case, the predictive check score is $0.49$ indicating that there is a very good agreement between the held-out events and the extracted PDF.  It is worth mentioning that the minimum (darkest) bin is 60\% of the maximum (brightest) bin, meaning that it is a smoothly changing PDF whose values do not go to close to zero in any region of the domain.  The reason for having such a PDF from where it is difficult to extract information from, it is precisely because this is already the signal region in the di-Higgs search.}}
    \label{posteriorpredictiva}
\end{figure}

The predictive check score provides a quantitative measurement on the compatibility between the replicate data set $X_{rep}$ and the held-out data, $X_{held}$.  However, depending on the specific features of the problem, one would have different expectations on how well the replicate data should mimic the true data.  For instance, in Ref.~\cite{blei-causal} a large matrix (with social data) should be expressed as a product of two matrices with considerable less dimensions, and they request predictive scores above a threshold of $0.10$.  In our case, however, we have smoothly changing PDFs in a bounded domain in $(m_{b\bar b}, m_{\gamma\gamma})$ in which neither the replicate data nor the true data PDFs take small values.  Observe that the maximum PDF variation in the domain is about $\sim~40\%$.  Therefore, we do expect more restrictive thresholds than in Ref.~\cite{blei-causal}.  In order to assess the predictive check score in the problem presented in this work, we have taken the replicate data set and artificially modified it in different combinations in some of its parameters and measure the predictive check score in each case while we also generate many different samples of $X_{held}$.  We present the results of this arbitrary exploration in Fig.~\ref{exploration}.  We can see that the true inference yields predictive scores of about $0.5 \pm 0.03$, which indicates good agreement between the replicate data and the true data.  We can also have a broad idea on the different sensitivity on the parameters, and in which parameters one would not have sensitivity if a threshold of about $\sim 0.45$ is required.  In particular, observe the little sensitivity to $\mu_{b\bar b}$ and $\sigma_{b\bar b}$, which is agreement to the corresponding discussion in the previous sub-section.

\begin{figure}[th]
    \centering
    \begin{minipage}[c]{\textwidth}
    \centering
    \includegraphics[width=0.65\textwidth]{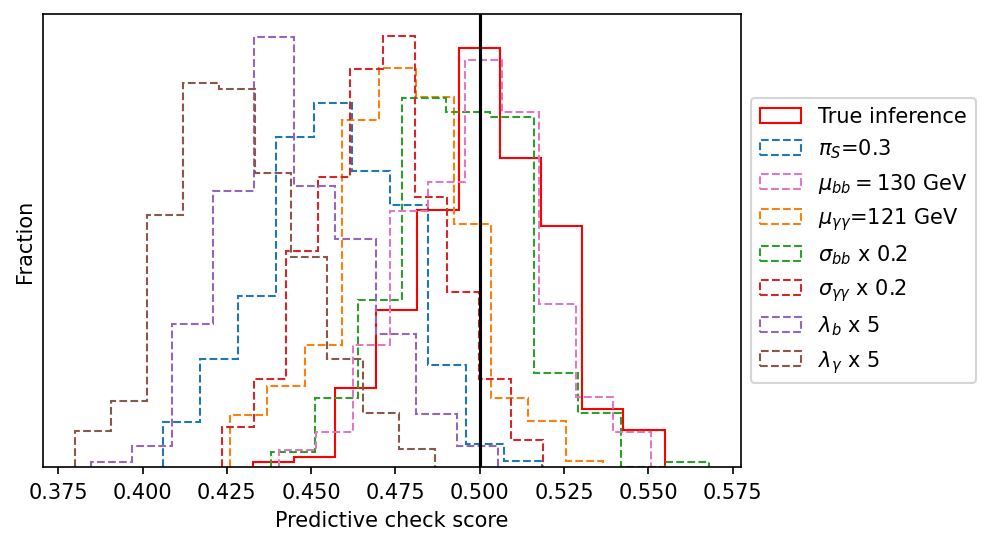}
    \end{minipage}
    \caption{\change{Predictive check score distributions for the true inference and for a few artificially shifted replicate data sets.  We indicate in the right panel which parameters have been artificially fixed/shifted for each case.  From the plot it can be recognized that the replicate data set from the inference process has a good agreement with the held-out data since it is centered around $0.5$, as expected.  We can also have a grasp on how much shift in the predictive check score is produced by different shifts in the parameters of the replicate data set PDF.  As expected, in these cases the predictive score goes below $0.5$, indicating the (injected) bias in the replicate PDF. From the plot it can be recognized that the data set and problem is not much sensitive to variations in $\mu_{b\bar b}$ and $\sigma_{b\bar b}$, this is in agreement with the discussion in Sect.~\ref{bi} and it is because of the very little variation that has the $m_{b\bar b}$ background in conjunction with large $\sigma_{b\bar b}$ and small signal fraction. }}
    \label{exploration}
\end{figure}

}
\section{Discussion}
There are a few matters about the presented method that it is interesting to discuss in more detail.  

Observe that one of the key features of the algorithm described above is to use some robust knowledge which has some unknown parameters which are then extracted from the data.  In our case, the robust knowledge is the shape of the invariant mass probability distributions for signal and background, but their exact parameters had to be learned from the data.  \change{Notice that the background parameters are not nuisance, but instead are also important in the process of extracting relevant features, such as for instance the signal fraction.}  The exact values of these parameters depend not only on the underlying physics at the matrix level, but also on many aspects which are usually not known in detail, such as hadronization, jet physics, and detector effects, among others. Therefore, the Bayesian inference process extracts these unknown from the data through the likelihood and the prior.  In the benchmark scenario in Section \ref{benchmark} we have used some assistance from a biased Montecarlo to construct the priors and, despite of this, we have shown to be able to extract good posteriors for the relevant parameters.

We have presented the results using some specific priors, as described in Fig.~\ref{fulldynesty}.  However, one should be cautious when determining the prior because the posterior has an explicit dependence on the former.  As a matter of fact, in our case in order to stress the power of the method we have used a wide prior for $\mu_{\gamma\gamma}$, and this could contribute to leaning the $\pi_S$ prediction towards lower values than its true value.  This is because having $\mu_{\gamma\gamma}$ too far away from its true value, where it does not find excess in events, is compensated by reducing the expected signal fraction.  In general, it is normal that a prior could systematically lean its posterior and therefore it is recommended to analyse and understand these effects.  In any case, we stress the benefits of using a prior \change{when there is prior information, since maximizing a likelihood is equivalent to assuming a flat prior, which may be missing some important and real prior information.}

We have presented an \change{over-}simplified scenario for di-Higgs at the LHC, however one could envisage other possible final states, such as for instance $ZZ$, $WW$, $t \bar t$, Leptoquark pair production and others in the same line.  That is, that the distribution of the final particles is different in signal and background, providing a leverage to distinguish signal from background taking advantage of the information encoded at the event-by-event level.  In some of these final states the decaying particles could be neutrinos, and hence the missing energy distribution would be a relevant variable.   If the missing energy has non-negligible contribution from more than one particle decay, then it is unlikely that the system would allow writing the conditional probability distribution as a direct product of each decaying particle, and hence the conditions described above would not be fulfilled.  Nevertheless, the problem should be studied in detail because some modified implementation could yield profitable results within a similar framework.  

The posterior probability distribution of the parameters is a tool that may have multiple uses.  For instance, in addition to providing the range of prediction for the parameters, one could build a tagger for the events.  In fact, given an event ($m_{b\bar b},\ m_{\gamma\gamma})$, the probability of belonging to a given class $k$ (signal or background) is
$$ p(z=k|m_{b\bar b},\ m_{\gamma\gamma}, X) = \int p(z=k|m_{b\bar b},\ m_{\gamma\gamma}, \theta) \, p(\theta|X) d\theta . $$
The last integral can be performed numerically using the sample of the posterior previously obtained.  This is similar to what has been done in Ref.~\cite{spannowsky} in another context.

Although the analysis in the previous Section is within an \change{over-}simplification \change{--and hence unrealistic--} of the di-Higgs scenario, the results in this work open the possibility to further explore this tool in a more realistic scenario.   In this sense it is interesting to briefly discuss the ATLAS \cite{atlas} and CMS \cite{cms} papers on di-Higgs in the channel $b\bar b \gamma \gamma$ in this context.  First, one should observe that backgrounds in which one of the pairs of $b\bar b$ or $\gamma\gamma$ is resonant and the other is not \change{($h\to b\bar b$ or $h\to \gamma\gamma$ plus non-resonant $\gamma\gamma$ or $b\bar b$, respectively, and others)}, can in principle be included \change{in the probabilistic model.  For instance, in $t\bar t h,\ h\to\gamma\gamma$ one would model the $\gamma\gamma$ as a Normal, and $b\bar b$ would have to be studied to see up to what extent it could be modeled as an exponential.  For other backgrounds such as for instance mis-tag/fake contributions, one should study up to what extent they could be included in the probabilistic model as having an exponentially decaying contribution and also study whether there are other statistic features in which they could be differentiated.  In any case, it is clear that the task of applying the presented techniques in a realistic di-Higgs analysis has to be investigated and studied in detail to explore its prospects.   Such an analysis would be very challenging, probably mainly because of the few expected signal events.}  Second, a few observations about the ATLAS and CMS current searches.  In ATLAS the search strategy \cite{atlas} consists in first applying a boosted decision tree (BDT) using many bottom and other variables, but avoiding including information about $m_{\gamma\gamma}$.  Then a maximum likelihood fit on $m_{\gamma\gamma}$ is performed to the events passing a given BDT score.  This workflow cannot fully see the above-discussed information encoded at the event-by-event level, and might be missing relevant information.  On the other hand, the CMS strategy \cite{cms} consists in training a network based on simulations in the $(m_{b\bar b}$, $m_{\gamma\gamma}$) plane, therefore the network does see the information encoded at the event-by-event level.  Both analyses have a reliance on the output of their Montecarlo simulation considerably greater than what is presented in this article.  Actually, it would be interesting to analyze (in this and/or another observable) at what point of a bias in a Montecarlo simulation it would become convenient to use an algorithm that learns mainly from the data rather than a network that learns from the biased Montecarlo.  

A few comments should be added about using this framework in a real di-Higgs search.  Since the prospects for di-Higgs measurements at the LHC are in the evidence region ($3\sigma$) combining all channels \cite{prospects, prospects2}, any innovation for a search strategy in this final state could be very important.  Moreover, the channel $hh\to b\bar b \tau \bar \tau$ is also another channel where this framework could be applied and still add more to the prospects of the di-Higgs measurement.

Another interesting point that raises the ATLAS and CMS papers is that they use a Crystall Ball function \cite{crystalball} to model the resonant decays.  This function is a slight modification to the Normal PDF and apparently fits better the corresponding invariant masses, at the price of using more variables than just $\mu$ and $\sigma$.  Within the framework presented in this article, there is a clear interplay between using more sophisticated and more precise PDFs, but with more parameters to adjust, or vice versa.  The interplay comes out because the more parameters there are to adjust, the more fuel (data) one needs to spend on their inference, and therefore the less precise is the posterior.  Since usually data is scarce, and therefore it has its own noisy fluctuations, then a balanced decision has to be taken on up to what level of sophistication one needs to describe the PDFs for the classes in the mixture model.

\change{
Finally a few words should be mentioned to justify proposing a Bayesian study in collider physics where the standard is a frequentist analysis.  In our experience, in most of the problems with enough data both frameworks correctly formulated end up giving approximately similar results hence, in this sense, any philosophical discrepancy gets diluted in the mentioned practical result.  However, there are at least two main reasons which in our understanding support performing analyses within a Bayesian statistics framework.   One of them is that the latest numerical and theoretical advancements  in Statistics and in the ML industry to obtain a posterior, or a MAP, from a complex model are outstanding (e.g. \cite{mlindustry, mlindustry2} and references therein) and can produce previously unthinkable results.  The second reason is Graphical Models, which provide a simple way to visualize, think, propose and understand sophisticated probabilistic models starting from basic principles, and many times with elementary PDFs.  The Bayesian framework is specially prepared to run and optimize inference departing from probabilistic models written as Graphical Models. These are some of the reasons which led to important achievements in the ML industry.  We pursue to apply these tools in collider physics in order to potentially improve current achievements.

}

\section{Conclusions}
\label{conclusions}

We have presented a\change{n application of a} statistical method to study signals consisting of a final state that comes from the decay of two or more particles.  We have shown that a Bayesian inference framework can efficiently distinguish such a signal from its background by taking advantage of the effects generated by their different distributions.  

We have shown in the article how some distributions for the particles in the final state acquire a non-negligible dependence when considering signal and background in the sample.  This dependence is a precise function of the parameters of the signal and background distributions, as well as their fraction in the sample.  The presented algorithm uses this dependence at the event-by-event level, and in the whole sample, to infer the posterior probability of the mentioned parameters.  To achieve this inference, we model the data as being created through a generative model using the corresponding hypothesis, and then infer the relevant latent variables using Bayesian techniques.  

To illustrate this idea we worked out a simplified search for $pp \to hh \to b \bar b \gamma \gamma$ at the LHC, where the two Higgs are the decaying particles in the signal final state.  In this case, the model has four parameters for the signal (two Normal distributions), two parameters for the background (two Exponential distributions), and the expected signal fraction in the sample.   We have  used {\tt Madgraph+Pythia+Delphes} to simulate LHC samples of $N=1000$ and $250$ events with signal fractions of 0\%, 5\% and 10\% and we have extracted the posterior over the parameters using dynamic nested sampling through the numerical library {\tt Dynesty}.  To assess the power of the algorithm we have used biased priors on the parameters and the method has learned from the data their posterior probabilities generally closer to the true values than the prior, and in all cases compatible with it.  This is a way to emulate a biased Montecarlo for a real LHC search, and since it has worked out correctly despite the injected bias, we argue that the algorithm reduces the impact of Montecarlo simulations on its results.  One of the main achievements of the proposed framework is its ability to learn from the data the expected signal fraction in the sample: Even though we set as prior a total agnostic uniform distribution on this parameter, the algorithm correctly extracts a posterior around its true value (see bottom panels in Figs.~\ref{1000} and \ref{250}). 

\change{As a validation of the inference procedure, we have performed a posterior predictive check on the results.  We have found a very good statistical agreement between a replicate data set sampled from the model using the posterior, and some held-out data not used in the inference process.  We have shown the usefulness of the posterior predictive to understand the sensitivity of the posterior in each parameter given the true data set.  For the di-Higgs simplified example we have found and justified the little sensitivity to the $m_{b\bar b}$ central value and standard deviation.}

We have discussed different aspects of the presented method, including how priors could affect the output; how this same framework could be used for other searches such as for instance $pp \to ZZ,\ tW,\ WW$, among others; how the posterior can be used to construct a signal-background tagger; how this tool could be applied to a real di-Higgs search;  we have discussed existing di-Higgs searches; and how using more sophisticated PDFs for signal and background has an interplay between adding parameters to increase modeling precision and having too wide and noisy posteriors because of the ratio between the number of parameters and the size of the data.

The proposed framework \change{pursues} to explore LHC searches from a different Bayesian perspective.  Work in this direction could provide a potential bias reduction and/or an enhancement in the sensitivity of existing LHC searches.

\section*{Acknowledgments}
Thanks to Manuel Szewc for fruitful discussions and for a careful reading of the manuscript.  Thanks to Leandro Da Rold for useful information on the di-Higgs backgrounds.  

\change{
\appendix
\section{Toy Model}
\label{toy}
We present the results of a toy model in which 500 data points are generated by a mixture of factorized Normal and Exponential distributions, as it is the likelihood used to extract the parameters.   This result is a test to verify that the inference procedure works correctly.   The script used to generate the data and the results are in Ref.~\cite{github}, and can be easily run with different parameters.

In this toy model example we have reduced the $m_{b\bar b}$ uncertainty ($\sigma_{b\bar b}$), increased the signal fraction ($\pi_S$), and increase the exponential parameters ($\lambda_{b,\gamma}$) in order to break existing degeneracies and have better contrast between the classes involved in the mixture.  Observe in Fig.~\ref{toymodel} how all the marginalized posteriors improve each prior in every parameter's true values, including $\sigma_{b\bar b}$ and $\sigma_{\gamma\gamma}$.

\begin{figure}[th]
    \centering
    \begin{minipage}[c]{\textwidth}
    \centering
    \includegraphics[width=0.32\textwidth]{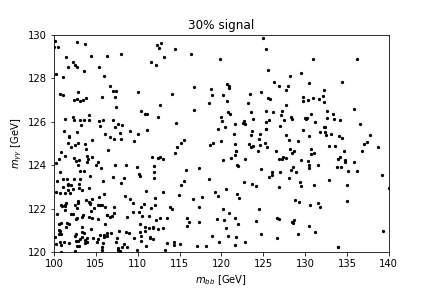}
    \includegraphics[width=0.32\textwidth]{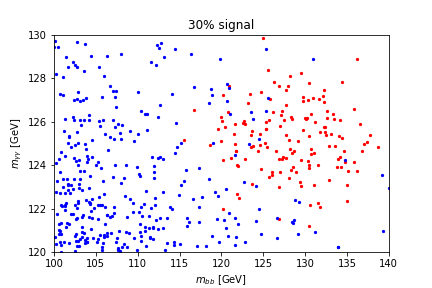} 
    
    \includegraphics[height=0.75\textheight]{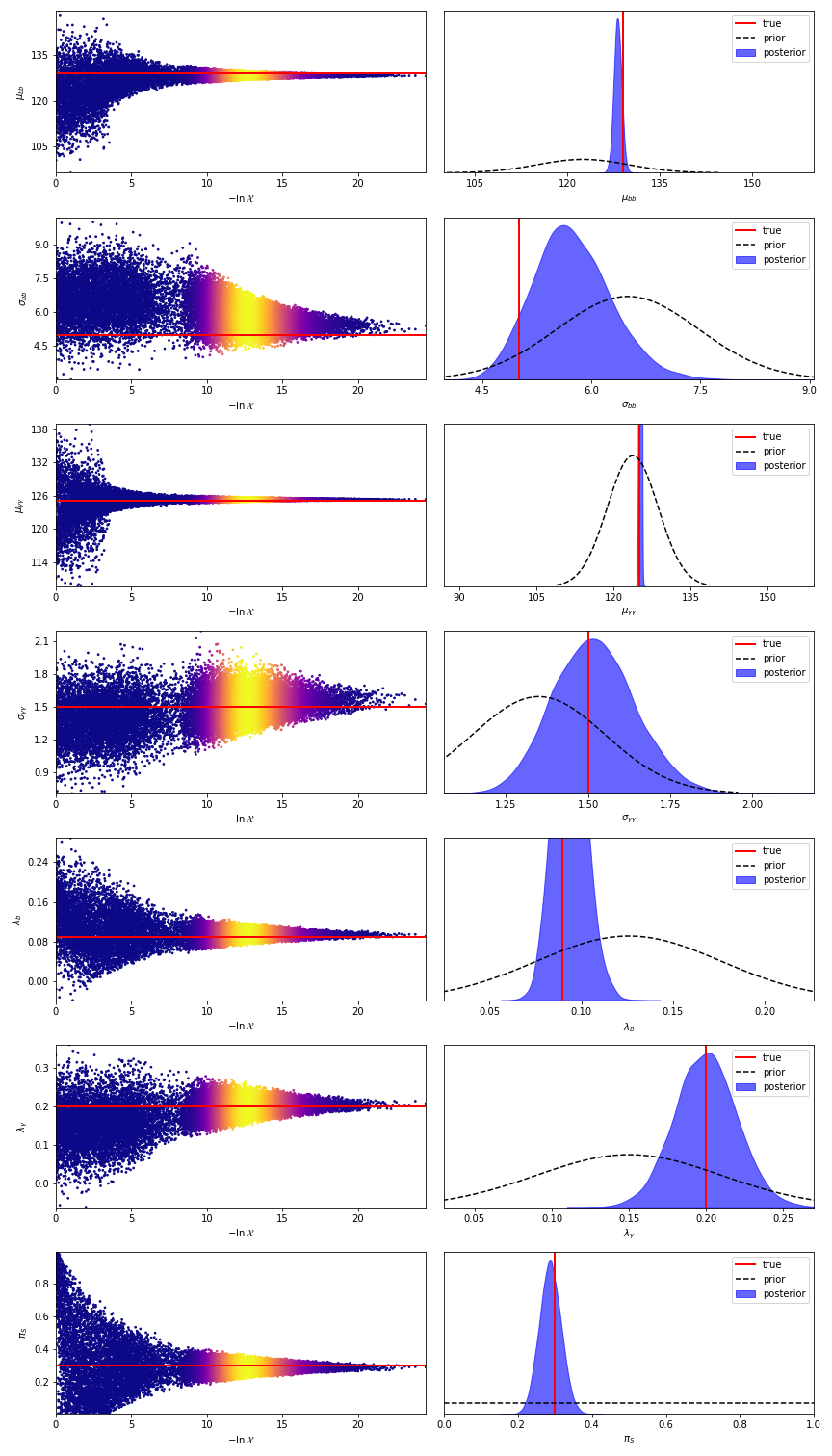}
    \end{minipage}
    \caption{\change{Bayesian inference run on 500 data points of fake data from a toy model.  We have created fake data which is generated exactly with 30\% of two factorized Normal distributions and 70\% of two factorized Exponential distributions.  We have extracted the parameters using the same model with which the data has been generated and obtained that the posterior improves the prior in all parameters. Upper row consists in the data distribution in the $m_{b\bar b}$--$m_{\gamma\gamma}$ plane without and with labels.  Main plot is the results of the inference process.        }} 
    \label{toymodel}
\end{figure}
}

\bibliography{biblio}

\end{document}